\documentclass[conference]{IEEEtran}%
\IEEEoverridecommandlockouts
\usepackage{cite}
\usepackage{footmisc}
\usepackage{amsmath,amssymb,amsfonts}
\usepackage{graphicx}
\usepackage{textcomp}
\usepackage{xcolor}
\usepackage{mathtools} %
\usepackage{todonotes} %
\usepackage[caption=false]{subfig} %
\usepackage[mathscr]{euscript} 
\usepackage{url} %
\usepackage{hyperref}
\usepackage{bm}
\usepackage{algpseudocode} 
\usepackage{algorithm}
\usepackage{xifthen}
\usepackage[export]{adjustbox}
\usepackage{textcomp}
\usepackage{array}   
\usepackage{svg}
\usepackage{enumitem}
\usepackage{xcolor}
\usepackage{booktabs}
\usepackage{enumitem}
\usepackage{todonotes}
\usepackage{xcolor}
\usepackage{tcolorbox}

\usepackage{wrapfig} 
\usepackage{colortbl}  
\newcommand{\mycomment}[1]{}
\NewDocumentCommand{\vect}{ O{} O{} m }{\mathbf{#3}\ifthenelse{\isempty{#1}}{}{^{(#1)}}\ifthenelse{\isempty{#2}}{}{_{#2}}}
\NewDocumentCommand{\mat}{ O{} O{} m }{\mathbf{#3}\ifthenelse{\isempty{#1}}{}{^{(#1)}}\ifthenelse{\isempty{#2}}{}{_{#2}}}
\NewDocumentCommand{\ten}{ O{} O{} m }{\pmb{\mathscr{#3}}\ifthenelse{\isempty{#1}}{}{^{(#1)}}\ifthenelse{\isempty{#2}}{}{_{#2}}}

\def\BibTeX{{\rm B\kern-.05em{\sc i\kern-.025em b}\kern-.08em
    T\kern-.1667em\lower.7ex\hbox{E}\kern-.125emX}}
\definecolor{mygreen}{rgb}{0,0.6,0}
\definecolor{mymauve}{rgb}{0.58,0,0.82}
\definecolor{mygray}{rgb}{0.5,0.5,0.5}
\definecolor{blue}{rgb}{0.0,0.0,1.0}
\definecolor{red}{rgb}{1.0,0.0,0.0}
\usepackage{listings}
\lstdefinelanguage{cypher2}{
    sensitive=true,
    morekeywords=[1]{MATCH, RETURN, WHERE, CONTAINS},
    morekeywords=[2]{PERSON, FRIEND,Document, Keyword, Affiliation, Country},
    morestring=[b]",
    morecomment=[l]{//},
    morecomment=[s]{/*}{*/},
    morecomment=[s]{--}{\ },
}
\lstdefinestyle{cypherstyle2}{
    language=cypher2,
    basicstyle=\footnotesize\ttfamily,
    keywordstyle=\color{blue}\bfseries, 
    keywordstyle=[2]\color{red}\bfseries,   
    commentstyle=\color{mygreen},
    stringstyle=\color{mymauve},
    numberstyle=\tiny\color{mygray},
    breaklines=true,
    showstringspaces=false,
    captionpos=b
}

\usepackage{listings}
\begin{document}
\title{Occupational Prompting Reveals Cultural Bias in Large Language Models}


\author{\IEEEauthorblockN{
Maksim E. Eren\IEEEauthorrefmark{1},
Andrea Brennen\IEEEauthorrefmark{2},
Ryan C. Barron\IEEEauthorrefmark{1},
and Eric Michalak\IEEEauthorrefmark{4}
}
\IEEEauthorblockA{
\IEEEauthorrefmark{1}Computational Intelligence \& Modeling, Los Alamos National Laboratory, Los Alamos, New Mexico, USA. \\
\IEEEauthorrefmark{2}IQT, USA. \\
\IEEEauthorrefmark{4}Advanced Research in Cyber Systems, Los Alamos National Laboratory, Los Alamos, New Mexico, USA.
}
\thanks{U.S. Government work not protected by U.S. copyright.}
}

\maketitle

\begin{abstract}
Social roles shape expectations, priorities, and judgments, yet it remains unclear how large language models (LLMs) associate occupational identities with broader cultural value patterns. Prior work used nationality-based cultural prompting to study how LLM responses to value-survey questions align with human cultural benchmarks. In this paper, we extend that framework by replacing cultural prompting with occupational prompting to examine how professional-role cues influence value-survey responses in open-weight LLMs. Using a survey-grounded evaluation pipeline based on questions from the Integrated Values Surveys, we project model responses into the two-dimensional Inglehart--Welzel cultural space. We prompt open-weight LLMs to answer questions under occupational identities such as accountant, teacher, engineer, and nurse, and then analyze how these occupation-conditioned responses are positioned on the cultural map. Our results show that when open-weight LLMs are prompted with occupations rather than national identities, their responses remain within a broadly Western-leaning region of the cultural map. However, different occupations introduce shifts within this region, producing distinct occupational skews. This indicates that occupational prompts are not treated as neutral role labels, but instead elicit structured value patterns. These findings extend survey-based evaluation of cultural bias beyond nationality-based prompting and provide a framework for studying how occupational personas shape value expression in LLMs.

\end{abstract}

\begin{IEEEkeywords}
Artificial Intelligence, LLM, Culture, Bias, Prompt Engineering
\end{IEEEkeywords}

\section{Introduction}
\label{sec:introduction}
Social roles influence how people interpret responsibility, authority, expertise, risk, and acceptable forms of judgment. Occupations, in particular, are not only descriptions of labor; they also carry assumptions about training, status, institutional norms, and interpersonal obligations. As large language models (LLMs) are increasingly used in analytical, professional, and decision-support workflows, it is important to understand whether they attach systematic cultural value orientations to occupational identities, and whether those associations shape model outputs in predictable ways.

A growing literature shows that LLMs are not culturally neutral. Prior studies have found that model outputs often reflect Western-leaning defaults, including value patterns associated with English-speaking or Western, Educated, Industrialized, Rich, and Democratic (WEIRD) societies \cite{johnson2022ghost, atari2023which, naous2024beer, AlKhamissi2024InvestigatingCA, Pawar2024SurveyOC, Zhou2025ShouldLB}. More broadly, recent work argues that cultural assumptions can enter not only through training data, but also through prompt design, evaluation design, and task framing 
\cite{navigli2023biases, Oh2025CultureIE, 10852463, kwok2024evaluating, Greco2026CulturallyGP, Eren2026PromptPF}. 
These concerns matter in settings where LLMs are used to summarize documents, support auditing, generate recommendations, or assist professional reasoning, because shifts in value expression can affect which trade-offs are emphasized and which judgments are presented as reasonable or legitimate \cite{KONIGSTORFER2022100043, Steen2023BiasIN, Yao2024SmartAS, Godbole2024LeveragingLL}. Recent work by Tao et al.\ \cite{pgae346} introduced a survey-grounded framework for measuring cultural bias by mapping LLM responses to value-survey questions into the Inglehart--Welzel cultural space \cite{inglehartwelzel2005}. Using the Integrated Values Surveys (IVS), which combine World Values Survey and European Values Study data \cite{haerpfer2022wvs, evs2022trend, ivs2023}, they showed that generic prompting produces a concentrated Western-skewed profile, while nationality-based prompting can move responses closer to country-level human benchmarks \cite{pgae346}. 

In this paper, we build on the survey-grounded framework of Tao et al.\ \cite{pgae346}, and study \emph{occupational prompting} to examine how professional-role cues influence survey responses in open-weight LLMs. We prompt models with occupational identities such as accountant, teacher, engineer, and nurse, ask them to answer the same value-survey questions, and project those responses into the same two-dimensional cultural space derived from human survey data. This allows us to test whether occupations induce systematic movement on the cultural map, whether different occupations occupy different regions, and whether broader occupation groupings defined by structural attributes such as the domain of the occupation exhibit patterns. Overall, we use occupational identities as probes for measuring how LLMs associate social roles with culturally inflected value profiles. This is important because occupational role descriptors are common in real prompting practice, and even when users do not explicitly invoke nationality or culture, these cues may still shape model behavior in meaningful ways. In that sense, occupational prompting provides a way to study how LLMs organize social-role information into latent value structure.

Using five open-weight LLMs and the IVS-based projection pipeline, we find that when models are conditioned on occupational identities, they generally remain within a larger Western-leaning region, but different occupations introduce shifts within it, producing distinct occupational skews on the cultural map. These skews suggest that models associate occupational domains with different value profiles along the two cultural axes, \textit{Survival vs. Self-Expression} and \textit{Traditional vs. Secular}, indicating that occupational prompts are not treated as neutral role labels but instead elicit structured patterns of value expression. In summary, our contributions are as follows:
\begin{enumerate}
    \item Extend survey-grounded evaluation of cultural bias in LLMs from nationality-based prompting to \emph{occupational prompting}, using professional identities to probe how role cues shape model responses.
    \item Project occupation-conditioned responses from open-weight LLMs into the Inglehart--Welzel cultural space and analyze how occupations and occupation groups are distributed within that benchmark framework.
    \item Study both individual occupations and metadata-based occupation groupings, enabling analysis of higher-level structure across domains and related occupational attributes.
    \item Compare multiple open-weight LLMs within the same survey-grounded pipeline to evaluate which occupational patterns are consistent across models and which are model-specific.
\end{enumerate}

\section{Related Works}
\label{sec:related_works}
Recent work has used survey instruments and persona prompting to examine the values, opinions, and social assumptions expressed by LLMs. Tao et al.~\cite{pgae346} introduce a survey-grounded cultural-alignment framework that projects LLM responses to World Values Survey (WVS) items into the Inglehart--Welzel cultural map and compares model placements against nationally representative benchmarks. Zhao et al.~\cite{Zhao2024WorldValuesBenchAL} similarly construct a large-scale value benchmark from WVS data to evaluate whether model responses align with human demographic and cultural distributions. Rozen et al.~\cite{ICLR2025_68fb4539} evaluate whether LLM-generated personas exhibit coherent value profiles, using the Schwartz theory of basic human values as a psychological benchmark. Rather than measuring only which values models endorse, they examine whether the relationships among values resemble human value structures, finding that generic prompting produces weak consistency while value-anchored prompting better matches human value correlations. More broadly, Tseng et al.~\cite{tseng-etal-2024-two} organize persona-based LLM research into role-playing and personalization, and Wang et al.~\cite{Wang2023RoleLLMBE} introduce RoleLLM and RoleBench to evaluate and improve character-level role-playing through role profiles, role prompting, and role-conditioned instruction tuning. Lutz et al.~\cite{Lutz2025ThePM} further show that the specific form of persona prompting matters, role-adoption formats and demographic priming strategies can change stereotyping, semantic diversity, and survey-response alignment. These studies establish that prompt-conditioned identities can substantially alter model behavior, but they primarily focus on national or demographic personas, character-role fidelity, prompt-format robustness, or value consistency. In contrast, our work uses occupations as the conditioning signal while retaining the same survey-grounded cultural map from \cite{pgae346}, allowing us to ask whether professional-role prompts induce systematic shifts in expressed cultural values.

Occupational bias has also been studied directly in LLM outputs, especially through gendered and demographic associations. Kotek et al.~\cite{Kotek2023GenderBA} evaluate gender stereotypes in occupational contexts by testing how models resolve ambiguous pronouns, finding that LLMs often reproduce stereotypical profession, gender associations and provide post-hoc rationalizations for those choices. Mirza et al.~\cite{Mirza2024EvaluatingGR} examine occupational and crime scenarios across LLMs by generating stories about professions and comparing inferred demographic distributions against U.S. labor and crime statistics. Jiang et al.~\cite{Jiang2025ExploringTO} extend occupational-bias analysis to Chinese LLMs by combining Chinese surnames with occupations and evaluating generated personal profiles for gender, age, regional, and educational stereotypes. These studies treat occupations as sites where models reveal demographic associations in generated text. Our study differs in both task and measurement, rather than asking which demographic attributes models assign to occupations, we prompt models to answer survey items from the standpoint of different occupations and then project those responses into the Inglehart--Welzel cultural space. This design makes it possible to measure occupation-conditioned value shifts on a shared cultural coordinate system, showing how occupational identities shape model-expressed values when the survey instrument and projection framework remain fixed.

\section{Methods}
\label{sec:methods}

We adapt the survey-grounded cultural-bias framework introduced in prior work \cite{pgae346}, but replace nationality-based prompting with occupational prompting in order to study how professional-role cues influence value expression in open-weight LLMs. Here we examine how occupations shift model responses to a fixed set of value-survey questions within the same benchmark cultural space of \cite{pgae346}.

\subsection{IVS benchmark space and cultural regions}

We construct the benchmark cultural space using the Integrated Values Surveys (IVS), which harmonize data from the World Values Survey (WVS) and the European Values Study (EVS) \cite{haerpfer2022wvs,evs2022trend,ivs2023}. Following prior work, we use the same ten survey items that underlie the Inglehart--Welzel cultural map \cite{inglehartwelzel2005,pgae346}. These items capture value dimensions related to happiness, social trust, authority, petition signing, religion, justifiability judgments, national pride, post-materialism, and child qualities. Survey responses are converted into numeric variables using the IVS/WVS/EVS coding guidance \cite{haerpfer2022wvs,evs2022trend,ivs2023}.

We fit Principal Component Analysis (PCA) on standardized respondent-level values and apply varimax rotation to obtain the canonical two-dimensional cultural space \cite{jolliffe2016pca,kaiser1958varimax}. As in prior work, the first two rotated components are interpreted as the \textit{Survival vs.\ Self-Expression} and \textit{Traditional vs.\ Secular} dimensions \cite{inglehartwelzel2005,pgae346}. We then apply the same linear rescaling used in earlier IVS-based replications:
\begin{equation}
PC1' = 1.81 \cdot PC1 + 0.38,
\end{equation}
\begin{equation}
PC2' = 1.61 \cdot PC2 - 0.01.
\end{equation}
Let $\boldsymbol{\mu}^{\mathrm{IVS}}_{\mathrm{raw}} \in \mathbb{R}^{10}$ and $\boldsymbol{\sigma}^{\mathrm{IVS}}_{\mathrm{raw}} \in \mathbb{R}^{10}$ denote the IVS means and standard deviations for the ten survey indicators, and let $W_{\mathrm{rot}} \in \mathbb{R}^{2 \times 10}$ denote the rotated PCA scoring matrix estimated from IVS data.

In addition to country-level coordinates, we retain the cultural \emph{Category} labels used in prior work, where countries are grouped into broader cultural regions on the benchmark map: African-Islamic, Catholic Europe, Confucian, English-Speaking, Latin America, Orthodox Europe, Protestant Europe, and West \& South Asia \cite{pgae346}. We represent each Category in two ways. For visualization, we estimate each region as a bivariate distribution in the PCA space using the empirical mean and covariance of the countries assigned to that Category; these are shown as soft covariance ellipses to illustrate dispersion and overlap in Figures \ref{fig:domain_single_panel_centroid} and \ref{fig:occupation_single_panel_centroid}. For assignment of LLM responses within this space, and distance-based analysis, we compute a Category centroid. Let $\mathcal{C}_r$ denote the set of countries assigned to Category $r$, and let $\boldsymbol{\nu}^{\mathrm{IVS}}_c \in \mathbb{R}^{2}$ denote the projected coordinate of country $c$. The centroid of Category $r$ is
\begin{equation}
\boldsymbol{\kappa}_r = \frac{1}{|\mathcal{C}_r|}\sum_{c \in \mathcal{C}_r} \boldsymbol{\nu}^{\mathrm{IVS}}_c.
\end{equation}
These centroids summarize the central tendency of the benchmark cultural regions and serve as the reference points for our assignment procedure.

\subsection{Occupation data, models, and prompting}

Our evaluation uses a curated set of 234 occupations together with structured metadata. We constructed this resource using ChatGPT Pro \cite{openai_chatgpt_pro_2026} as a dataset-curation tool, generating an initial occupation inventory and associated metadata fields for each occupation. This use of an LLM is motivated by recent work showing that LLMs can support practical dataset construction by generating synthetic examples or approximate annotations, thereby reducing the cost of curation and expanding coverage across categories and scenarios that may be difficult to collect manually \cite{wang2023selfinstruct,bansal2023llmannotators,long2024survey}. The generated entries were designed to span professional, technical, service, scientific, creative, public-sector, and skilled-trade roles, and each occupation was annotated with metadata describing broader occupational structure, including sector, domain, education field, and education level. Following prior work, we view such LLM-curated data as useful when paired with review and basic quality-control steps such as filtering, deduplication, and targeted sampling \cite{wang2023selfinstruct,bansal2023llmannotators}. At the same time, because LLM-generated inventories and annotations may reflect model-specific biases and need not reproduce real-world occupational or demographic distributions, we treat this dataset as a pragmatic analytical support rather than a substitute for ground-truth observational data \cite{Bisbee2024,long2024survey}. In the analyses reported here, we focus on individual occupations and on \textit{domain}-level groupings, where a \textit{domain} denotes a mid-level occupational field that groups substantively related jobs, such as \textit{data and AI}, \textit{accounting and audit}, \textit{performing arts}, \textit{medical practice}, and \textit{software engineering}.

We evaluate open-weight LLMs spanning different architectures, scales, and training regimes: Llama 3.3 (70B), Llama 4 (16$\times$17B), Gemma 3 (27B), GPT-OSS (20B), and GPT-OSS (120B) \cite{llama3_2024,llama4_2024,gemma3_2024,gptoss_2024}. For each of the ten IVS items, we prompt each model using the original survey question text together with strict response-format instructions so that outputs can be mapped deterministically to the corresponding numeric variables, following the same general procedure used in prior survey-grounded work \cite{pgae346}. The key change is in the identity prefix: instead of nationality-based prompting, we prepend each survey question with an occupational identity statement. To reduce sensitivity to small wording changes, we preserve the same respondent-descriptor variants used in earlier work (for example, "average human being", "typical human being", "average person", etc.) \cite{pgae346}. The use of this prompting strategy does not reflect the authors' belief that any such "average human being" exists; rather, it is employed here as an established elicitation technique, documented in the literature, for surfacing latent biases in LLM outputs. For occupation-conditioned prompting, each descriptor is combined with an occupation label. For example, an occupationally conditioned prompt takes the form ``You are an average human being working as an accountant responding to the following survey question.'' The remainder of the prompt contains the original survey question and the constrained response format. Box~\ref{box:example_prompt_occ} shows an example of the generic and occupation-conditioned versions of one IVS item.

\begin{tcolorbox}[title={Example prompting (generic vs.\ occupational).\\
A008 (Feeling of Happiness) Question from IVS}, colback=white, colframe=black]
\small
\textbf{Generic (no occupational prompting):}\\
\texttt{Question: Taking all things together, rate how happy you would say you are.}\\
\texttt{Please use a scale from 1 to 4, where 1 is Very happy, 2 is Quite happy,}\\
\texttt{3 is Not very happy, 4 is Not at all happy.}\\
\texttt{You can only respond with a score number based on the scale provided and please do not give reasons.}\\
\texttt{Your score number:}

\vspace{0.6em}
\textbf{Occupational prompting (occupation-conditioned):}\\
\textcolor{red}{\texttt{You are an average human being working as an accountant responding to the following survey question.}}\\
\texttt{Question: Taking all things together, rate how happy you would say you are.}\\
\texttt{Please use a scale from 1 to 4, where 1 is Very happy, 2 is Quite happy,}\\
\texttt{3 is Not very happy, 4 is Not at all happy.}\\
\texttt{You can only respond with a score number based on the scale provided and please do not give reasons.}\\
\texttt{Your score number:}
\label{box:example_prompt_occ}
\end{tcolorbox}

Let $\mathbf{x}_{m,o,v} \in \mathbb{R}^{10}$ denote the coded response vector produced by model $m$ under occupation condition $o$ and descriptor variant $v$. We standardize and project this vector into the IVS benchmark space using IVS-derived moments and the rotated PCA scoring map:
\begin{equation}
\mathbf{z}_{m,o,v} = \left(\mathbf{x}_{m,o,v} - \boldsymbol{\mu}^{\mathrm{IVS}}_{\mathrm{raw}}\right) \oslash \boldsymbol{\sigma}^{\mathrm{IVS}}_{\mathrm{raw}},
\end{equation}
\begin{equation}
\mathbf{s}_{m,o,v} = W_{\mathrm{rot}} \mathbf{z}_{m,o,v},
\end{equation}
where $\oslash$ denotes elementwise division. We then apply the IVS rescaling to obtain a two-dimensional coordinate $\boldsymbol{\pi}_{m,o,v} \in \mathbb{R}^{2}$. To reduce wording sensitivity, we average across the respondent-descriptor variants:
\begin{equation}
\boldsymbol{\mu}_{m,o} = \frac{1}{|V|}\sum_{v \in V} \boldsymbol{\pi}_{m,o,v}.
\end{equation}
Here, $\boldsymbol{\mu}_{m,o}$ is the final occupation-conditioned coordinate for model $m$ and occupation $o$.

\subsection{Centroid-based assignment and analysis}

Each occupation is represented by its final projected coordinate $\boldsymbol{\mu}_{m,o}$ in the IVS cultural space. In the visualizations, we also plot these occupation-level points directly to show how raw occupation-conditioned responses disperse around the higher-level reference structure. For domain-level analysis, we compute a centroid by averaging the coordinates of all occupations assigned to a given domain. Let $\mathcal{O}_a$ denote the set of occupations in domain $a$. The domain centroid for model $m$ is
\begin{equation}
\bar{\boldsymbol{\mu}}_{m,a} = \frac{1}{|\mathcal{O}_a|}\sum_{o \in \mathcal{O}_a} \boldsymbol{\mu}_{m,o}.
\end{equation}

We assign both occupation points and domain centroids to benchmark cultural Categories using nearest-centroid matching in Euclidean distance. For an occupation or domain point $\mathbf{q} \in \mathbb{R}^{2}$, the assigned cultural Category is
\begin{equation}
\hat{r}(\mathbf{q}) = \arg\min_{r} \left\| \mathbf{q} - \boldsymbol{\kappa}_r \right\|_2,
\end{equation}
where $\boldsymbol{\kappa}_r$ is the centroid of benchmark Category $r$. This produces a partition of occupations and domains over the cultural space aligned with the empirical region structure. Because the assignment is based on Category centroids rather than individual countries, it emphasizes region-level structure rather than local country-specific variation. In addition to direct assignment, in Figure \ref{fig:top5_by_category_model_strips}, we analyze which occupations are most associated with each benchmark cultural Category by ranking them according to Euclidean distance to that Category centroid. For Category $r$, the association score for occupation $o$ under model $m$ is
\begin{equation}
d_r(m,o) = \left\| \boldsymbol{\mu}_{m,o} - \boldsymbol{\kappa}_r \right\|_2.
\end{equation}
We then rank occupations for each Category $r$ in ascending order of $d_r(m,o)$ and report the top occupations per region. This identifies which occupational prompts lie closest to each benchmark cultural Category in the IVS cultural space. IVS map provides a common reference frame for comparing how models organize occupational identities relative to established cultural dimensions.
 
 \section{Results}
\label{sec:results}
\begin{figure*}[t]
  \centering
  \includegraphics[
    width=\textwidth,
    height=0.82\textheight,
    keepaspectratio
  ]{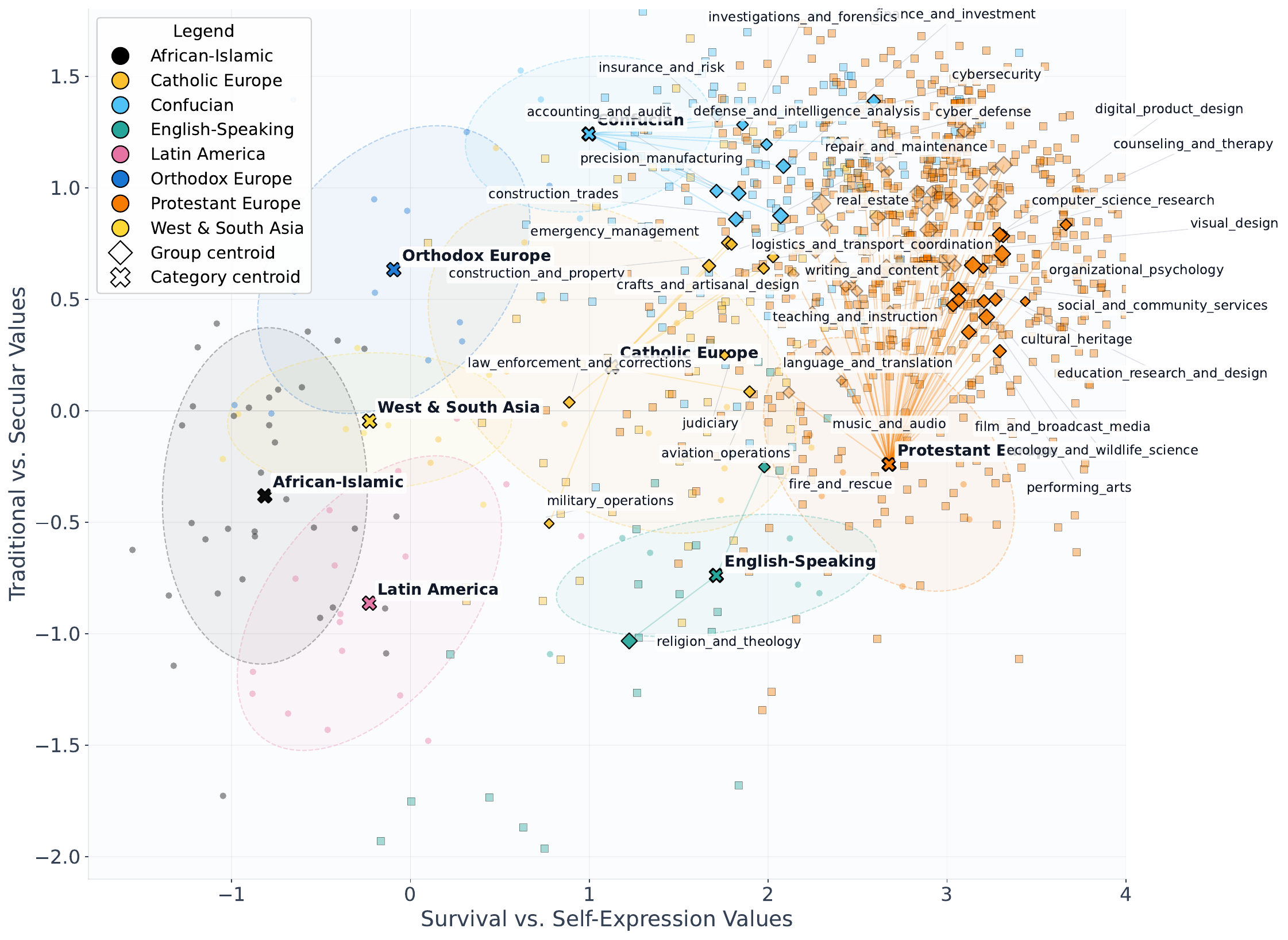}
  \caption{Domain-level occupation-conditioned responses projected into the IVS benchmark cultural space after averaging coordinates across the five open-weight LLMs. Country/territory points provide the background map, squares show occupation-conditioned responses colored by nearest cultural Category assignment, highlighted markers indicate Category centroids, and ellipses summarize within-Category country dispersion. Annotated labels indicate the top-ranked occupational domains assigned to each cultural Category.}
  \label{fig:domain_single_panel_centroid}
\end{figure*}

Figure~\ref{fig:domain_single_panel_centroid} shows domain-level occupation-conditioned responses projected into the IVS benchmark cultural space after averaging coordinates across the five open-weight LLMs. The domain centroids do not span the full country/territory distribution. Instead, most domains remain on the self-expression side of the map, with a dense concentration around Protestant Europe and neighboring Western or near-Western regions. This indicates that occupational prompting introduces variation within the cultural map, but does not eliminate the broader Western-leaning prior observed in prior survey-grounded cultural-bias work. The main structure is therefore not a full redistribution across global cultural regions, but a patterned displacement within a comparatively narrow portion of the IVS space.

Within this constrained region, however, the domain-level placements show interpretable occupational differences. Domains such as \textit{digital product design}, \textit{computer science research}, \textit{visual design}, \textit{counseling and therapy}, \textit{organizational psychology}, \textit{social and community services}, and \textit{education research and design} lie far to the self-expression side and cluster near the Protestant Europe region. By contrast, \textit{accounting and audit}, \textit{insurance and risk}, \textit{defense and intelligence analysis}, \textit{cybersecurity}, and \textit{cyber defense} shift upward toward the secular side of the map and closer to the Confucian region. \textit{Construction}, \textit{repair}, \textit{logistics}, \textit{emergency management}, and \textit{law enforcement} domains occupy a more central position, closer to the boundaries among Catholic Europe, West \& South Asia, and Orthodox Europe. \textit{Religion and theology} is distinctive because it moves downward toward the more traditional side of the space and lies closest to the English-Speaking region. These patterns suggest that models associate occupational domains with different value profiles along both axes: technical, financial, risk, and security domains shift toward more secular and relatively less self-expressive coordinates, while creative, social, educational, and design-oriented domains remain closer to the high self-expression Western cluster.

\begin{figure*}[t]
  \centering
  \includegraphics[
    width=\textwidth,
    height=0.82\textheight,
    keepaspectratio
  ]{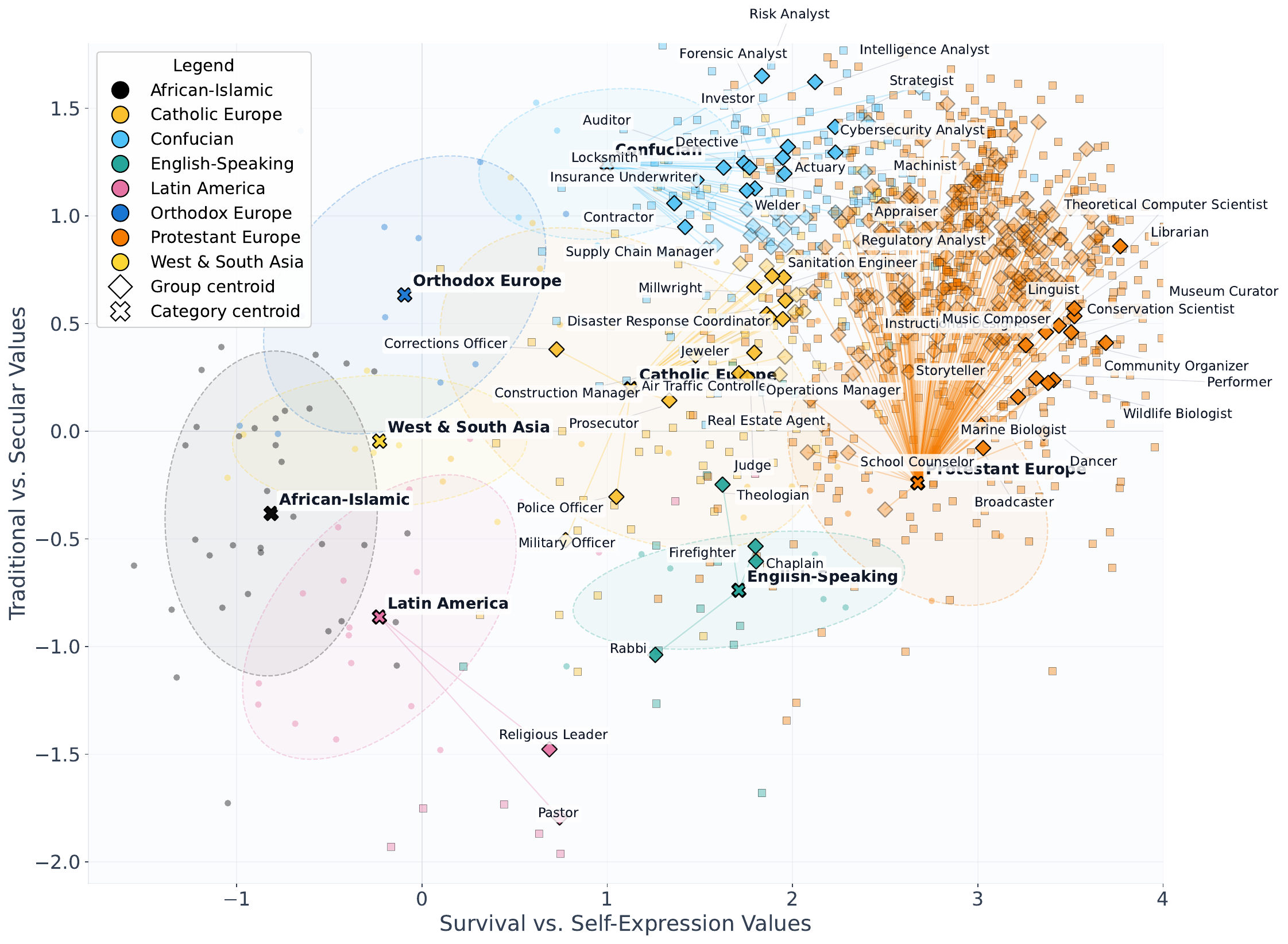}
  \caption{Occupation-level placements in the IVS benchmark cultural space (Survival vs.\ Self-Expression; Traditional vs.\ Secular values) after averaging coordinates across the five open-weight LLMs. The background shows country/territory points from the IVS map, while occupation-conditioned responses are shown as squares and colored by their nearest cultural Category assignment. Category centroids are highlighted, and covariance ellipses summarize the dispersion of countries within each benchmark cultural region. Annotated labels show the top-ranked occupations assigned to each cultural Category, revealing which occupational prompts lie closest to each region centroid under the projection procedure.}
  \label{fig:occupation_single_panel_centroid}
\end{figure*}

Figure~\ref{fig:occupation_single_panel_centroid} provides a more granular view by plotting individual occupations rather than domain centroids. The occupation-level placements show greater dispersion, indicating that domain aggregation smooths over meaningful role-specific variation. Several occupations linked to finance, risk, security, and analytic control occupy the upper portion of the map, including \textit{Risk Analyst}, \textit{Investor}, \textit{Auditor}, \textit{Insurance Underwriter}, \textit{Forensic Analyst}, \textit{Intelligence Analyst}, \textit{Cybersecurity Analyst}, \textit{Strategist}, and \textit{Actuary}. These roles are often assigned near the Confucian region, consistent with the domain-level pattern for accounting, insurance, intelligence, and cyber-related fields. In contrast, occupations such as \textit{Theoretical Computer Scientist}, \textit{Librarian}, \textit{Museum Curator}, \textit{Conservation Scientist}, \textit{Community Organizer}, \textit{Wildlife Biologist}, and \textit{Dancer} appear farther to the self-expression side and closer to Protestant Europe.

The occupation-level map also shows that authority- and institution-oriented occupations move away from the densest Protestant Europe cluster. \textit{Police Officer}, \textit{Military Officer}, \textit{Corrections Officer}, \textit{Judge}, \textit{Prosecutor}, \textit{Construction Manager}, and \textit{Disaster Response Coordinator} appear closer to the central or left-of-center part of the map, near Catholic Europe, West \& South Asia, or Orthodox Europe boundaries. Religious occupations form another distinct pattern: \textit{Pastor} and \textit{Religious Leader} shift strongly downward toward the Latin America region, while \textit{Rabbi} and \textit{Chaplain} remain closer to English-Speaking coordinates. This indicates that occupation-conditioned prompting can produce role-specific movements that are masked at the domain level. However, even these more extreme occupations generally do not reach the African-Islamic or West \& South Asia centroids closely; rather, they move in those directions from within a still-compressed model-response region.

\begin{figure*}[htb]
  \centering
  \includegraphics[width=\textwidth,height=0.88\textheight,keepaspectratio]{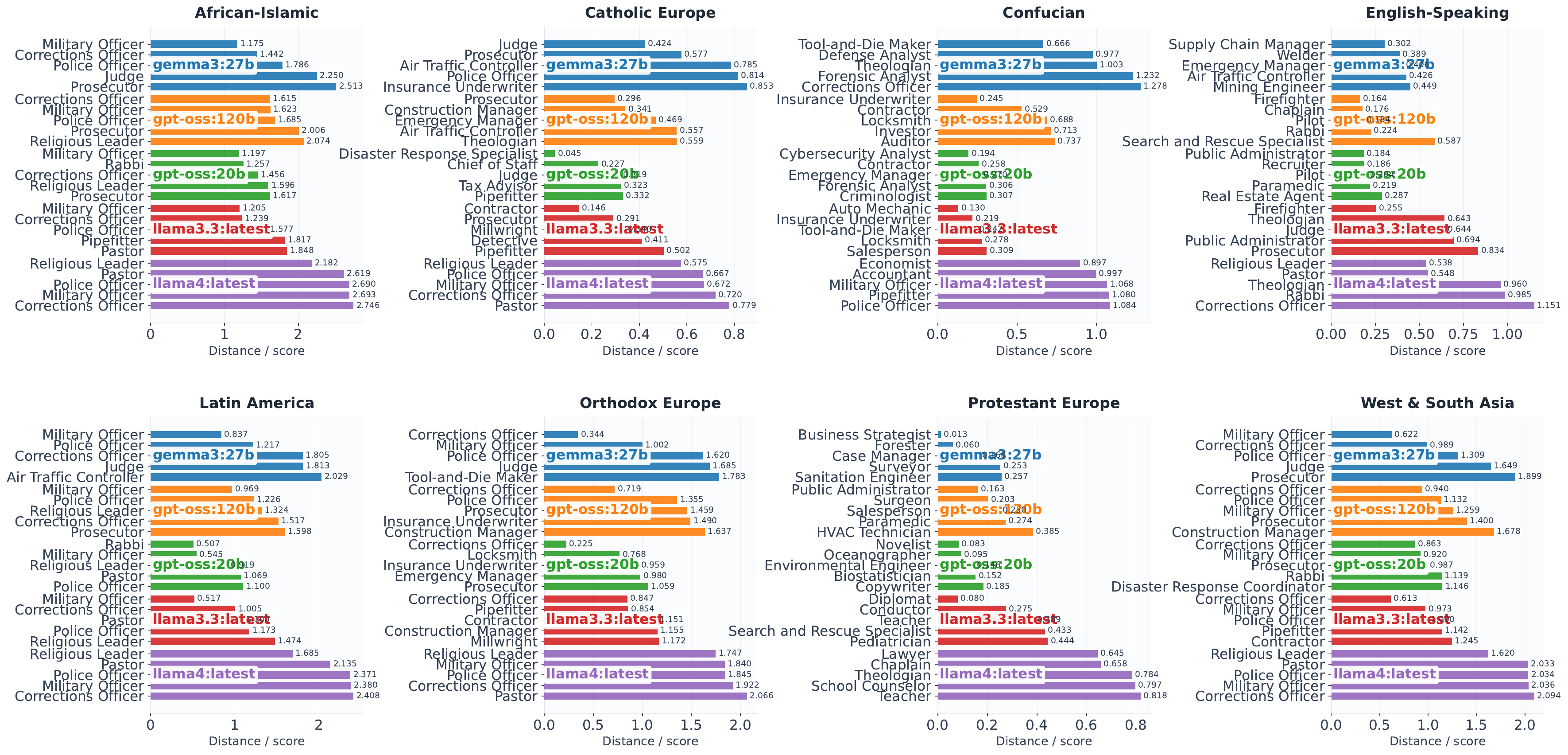}
  \caption{Top occupation--Category associations by model. For each benchmark cultural Category, we rank occupation-conditioned responses by Euclidean distance to the corresponding Category centroid in the IVS cultural space. Lower distance indicates that an occupation prompt lies closer to that cultural-region centroid. Results are separated by model, showing which occupations each open-weight LLM places nearest to African-Islamic, Catholic Europe, Confucian, English-Speaking, Latin America, Orthodox Europe, Protestant Europe, and West \& South Asia regions. This view complements the map-based visualizations by summarizing the nearest occupation prompts for each cultural Category rather than showing all projected points.}
\label{fig:top5_by_category_model_strips}
\end{figure*}

Figure~\ref{fig:top5_by_category_model_strips} further shows that the benchmark cultural Categories are not equally reachable through occupational prompting. Protestant Europe has the smallest nearest-centroid distances across models and also the most diverse set of nearest occupations, including strategic, environmental, scientific, public-service, educational, creative, and care-oriented roles. Catholic Europe, Confucian, and English-Speaking are also relatively close for several models, but their nearest occupations are more specialized: Catholic Europe is associated with legal, emergency-management, construction, and public-order roles; Confucian is associated with technical, financial, risk, security, and skilled-trade roles; and English-Speaking is associated with emergency response, transport, public administration, religious service, and real-estate/service occupations.

By contrast, African-Islamic, Latin America, Orthodox Europe, and West \& South Asia are reached through a narrower and more repetitive set of occupations. Across models, the nearest occupations for these Categories repeatedly include \textit{Military Officer}, \textit{Police Officer}, \textit{Corrections Officer}, \textit{Judge}, \textit{Prosecutor}, \textit{Pipefitter}, \textit{Contractor}, \textit{Religious Leader}, and \textit{Pastor}. This pattern suggests that, when occupation-conditioned responses move away from the high self-expression Western cluster, they often do so through authority-, public-order-, trade-, and religion-linked occupational cues rather than through a broad range of professional identities. The result is not symmetric cultural coverage of the IVS map, but a structured set of occupational pathways into some regions and a much narrower set of approximations for others.

The model-separated rankings show that model differences are not limited to distance magnitude; different models also use different occupational pathways to approach the same cultural Category centroid. For Orthodox Europe, Gemma~3, GPT-OSS~120B, and GPT-OSS~20B mostly select correctional, policing, legal, insurance, or emergency-adjacent occupations, whereas Llama~3.3 shifts toward trades and construction roles such as \textit{Pipefitter}, \textit{Contractor}, \textit{Construction Manager}, and \textit{Millwright}, and Llama~4 adds a stronger religion-linked pattern through \textit{Religious Leader} and \textit{Pastor}. Protestant Europe differs because it has both smaller distances and broader occupational diversity: Gemma~3 selects managerial, environmental, and service roles; GPT-OSS~20B selects scientific and creative roles; Llama~3.3 selects diplomacy, education, arts, and care-related roles; and Llama~4 emphasizes law, counseling, education, and religious professions. Confucian assignments are more concentrated around technical, risk, finance, and security work, but the pathway still varies by model: GPT-OSS~120B emphasizes insurance, investment, and audit roles; GPT-OSS~20B emphasizes cybersecurity, emergency management, forensics, and criminology; Llama~3.3 emphasizes skilled technical and service roles; Gemma~3 combines technical, defense, forensic, and correctional occupations; and Llama~4 shifts toward economics, accounting, military, trades, and policing. Thus, model family affects not only how close occupation-conditioned responses move toward each centroid, but also which occupational semantics are used to approximate the same benchmark cultural region.

\section{Conclusion}
\label{sec:conclusion}
We extended survey-grounded evaluation of cultural bias in large language models from nationality-based prompting to occupational prompting. Using the IVS-based cultural space, we showed that open-weight LLMs retain the broad Western-skewed pattern observed under generic prompting, while occupational identities introduce shifts within that larger region. These results indicate that occupational prompts are not treated as neutral role labels, but instead elicit structured value patterns.

Finally, this study has several limitations. First, the occupation set and metadata were curated with LLM assistance, making the dataset a practical analytical framework rather than a substitute for real occupational survey data. Second, our evaluation is based on short-form, forced-choice survey responses, which may not fully reflect how occupational cues influence longer-form reasoning or downstream task behavior. Third, the projections should not be interpreted as measuring the true cultural values of real-world professions; they instead capture how models organize occupational identities relative to a benchmark cultural map. A natural direction for future work is to test whether cultural prompting when combined with occupational prompting affects downstream task quality in applied settings, such as programming, analytical writing, recommendation, or domain-specific decision support.

\section*{Acknowledgment}
This manuscript has been approved for unlimited release and has been assigned LA-UR-26-23832. The funding for this paper was provided by Los Alamos National Laboratory (LANL). LANL is operated by Triad National Security, LLC, for the National Nuclear Security Administration of the U.S. Department of Energy (Contract No. 89233218CNA000001).

\bibliographystyle{IEEEtran}
\bibliography{References}

\vspace{12pt}

\end{document}